\documentclass[aps,pra,twocolumn,superscriptaddress,longbibliography]{revtex4-1}
\usepackage{amsmath}
\usepackage{bm}
\usepackage{graphicx}
\usepackage[breaklinks=true]{hyperref}
\hypersetup{colorlinks=true,citecolor=magenta,linkcolor=magenta,urlcolor=magenta}
\usepackage{breakcites}

\begin{document}

%\preprint{}

\title{Bounds for nonadiabatic transitions}


\author{Takuya Hatomura}
\email[]{takuya.hatomura.ub@hco.ntt.co.jp}
%\thanks{}
\affiliation{NTT Basic Research Laboratories, NTT Corporation, Kanagawa 243-0198, Japan}
\affiliation{NTT Communication Science Laboratories, NTT Corporation, Kanagawa 243-0198, Japan}

\author{Go Kato}
\affiliation{NTT Communication Science Laboratories, NTT Corporation, Kanagawa 243-0198, Japan}
\affiliation{NTT Research Center for Theoretical Quantum Physics, NTT Corporation, Kanagawa 243-0198, Japan}


\date{\today}

\begin{abstract}
We discuss bounds for nonadiabatic transitions from the viewpoints of the adiabatic perturbation theory and the quantum speed limit. 
We show that the amount of nonadiabatic transitions from the $n$th level to the $m$th level is bounded by a function of the quantum geometric tensor for the $m$th level. 
We analyze this bound from the viewpoint of the adiabatic perturbation theory. 
In addition, this bound and the viewpoint of the quantum speed limit suggest nontrivial relationship between the dynamical transformation and the adiabatic transformation. 
We also derive a universal bound for any nonadiabatic transition. 
This bound is written in terms of the counterdiabatic Hamiltonian. 
\end{abstract}

%\keywords{}

\maketitle


\section{Introduction}
Nonadiabatic transitions, which are transitions between different energy eigenstates due to fast change of Hamiltonians in time, have been studied since the dawn of quantum mechanics~\cite{Landau1932,Zener1932,Stueckelberg1932,Majorana1932,Shevchenko2010}. 
Generally, energy gap closing leads dynamics to critical slowing down, i.e., around critical points states are frozen and cannot follow change of energy eigenstates, and thus transitions between different energy eigenstates take place~\cite{Kibble1976,Kibble1980,Zurek1985,Zurek1996,delCampo2014}.

Evaluation of the amount of nonadiabatic transitions is of interest. 
Besides the traditional Landau-Zener formula~\cite{Landau1932,Zener1932,Stueckelberg1932,Majorana1932,Shevchenko2010} and the Kibble-Zurek formula~\cite{Kibble1976,Kibble1980,Zurek1985,Zurek1996,delCampo2014}, we can approximately calculate the amount of nonadiabatic transitions by using the adiabatic perturbation theory~\cite{Polkovnikov2005,Polkovnikov2011,Kolodrubetz2017}. 
It was shown that the amount of nonadiabatic transitions after a quantum quench is expressed by the adiabatic gauge potential, and as the result by the quantum geometric tensor.

Not only approximate evaluation but also bounds for the amount of nonadiabatic transitions are also of great interest. 
As predicted from the adiabatic theorem~\cite{Kato1950,Avron1987,Jansen2007,Lidar2009}, it is known that bounds are related to (minimal) energy gaps and change of energy eigenstates~\cite{Zhang2019}. 
Quantum speed limits, which give typical time scale to achieve quantum processes~\cite{Mandelstam1945,Pfeifer1995,Margolus1998,Deffner2017}, can be also used to quantify deviation of a dynamical state from an adiabatic state~\cite{Lychkovskiy2017,Suzuki2019}.

Recently, suppression of nonadiabatic transitions has been paid much attention due to development of adiabatic quantum computation~\cite{Farhi2000,Farhi2001,Albash2018} and quantum annealing~\cite{Kadowaki1998,Johnson2011}. 
In particular, finite time processes are important not only to achieve fast operation but also to avoid decoherence. 
Quantum adiabatic brachistochrone is a promising approach for parameter scheduling to reduce the amount of nonadiabatic transitions~\cite{Rezakhani2009,Rezakhani2010}. 
Use of shortcuts to adiabaticity is also remarkable strategy to suppress nonadiabatic transitions~\cite{Demirplak2003,Berry2009,Chen2010,Guery-Odelin2019}.

In this paper, we discuss bounds for nonadiabatic transitions in general quantum dynamics from the viewpoints of the adiabatic perturbation theory and the quantum speed limit. 
We find that the amount of nonadiabatic transitions is bounded by a function of the quantum geometric tensor, and thus our result is clearly consistent with an approximate calculation by the adiabatic perturbation theory~\cite{Polkovnikov2005,Polkovnikov2011,Kolodrubetz2017}. 
Note that this bound is also related to quantum adiabatic brachistochrone~\cite{Rezakhani2009,Rezakhani2010}. 
We also derive a universal bound for any nonadiabatic transition. 
This bound is written in terms of the counterdiabatic Hamiltonian, and thus we can also find the relationship to shortcuts to adiabaticity~\cite{Demirplak2003,Berry2009,Chen2010,Guery-Odelin2019}. 
We also analyze our results from the viewpoint of the quantum speed limit~\cite{Mandelstam1945,Pfeifer1995,Margolus1998,Deffner2017}. 
Our finding and the viewpoint of the quantum speed limit provide us with nontrivial relationship between the adiabatic transformation and the dynamical transformation.

\section{Bounds for nonadiabatic transitions}
We consider a quantum system described by the time-dependent Hamiltonian
\begin{equation}
\hat{H}(\lambda_t)=\sum_nE_n(\lambda_t)\hat{P}_n(\lambda_t),
\label{Eq.ham}
\end{equation}
where $E_n(\lambda_t)$ is the energy eigenvalue and $\hat{P}_n(\lambda_t)$ is the associated projection operator. 
Here, this Hamiltonian depends on time through the time-dependent parameter $\lambda_t=\{\lambda_t^{(1)},\lambda_t^{(2)},\cdots\}$. 
Dynamics of this system is generated by the time evolution operator $\hat{U}_D(t)$ satisfying
\begin{equation}
i\hbar\dot{\hat{U}}_D(t)=\hat{H}(\lambda_t)\hat{U}_D(t),\quad\hat{U}_D(0)=1. 
\label{Eq.Schrodinger.eq}
\end{equation}
Throughout this paper, the dot symbol represents time derivative.

Time evolution of the projection operators is isometrically generated by the adiabatic transformation $\hat{U}_A(t)$ as
\begin{equation}
\hat{P}_n(\lambda_t)=\hat{U}_A(t)\hat{P}_n(\lambda_0)\hat{U}_A^\dag(t), 
\label{Eq.ad.trans}
\end{equation}
for all $n$. 
According to the theory of shortcuts to adiabaticity~\cite{Guery-Odelin2019}, the adiabatic transformation satisfies
\begin{equation}
i\hbar\dot{\hat{U}}_A(t)=[\hat{H}(\lambda_t)+\hat{H}_\mathrm{cd}(t)]\hat{U}_A(t),\quad\hat{U}_A(0)=1, 
\label{Eq.tevo.adiabatic}
\end{equation}
where $\hat{H}_\mathrm{cd}(t)$ is the counterdiabatic Hamiltonian given by
\begin{equation}
\hat{H}_\mathrm{cd}(t)=\frac{i\hbar}{2}\sum_n[\dot{\hat{P}}_n(\lambda_t),\hat{P}_n(\lambda_t)]. 
\label{Eq.cd.ham}
\end{equation}
One may use the adiabatic Hamiltonian~\cite{Avron1987} (the reduced counterdiabatic Hamiltonian~\cite{Takahashi2013})
\begin{equation}
\hat{H}_\mathrm{cd}^{(n)}(t)=i\hbar[\dot{\hat{P}}_n(\lambda_t),\hat{P}_n(\lambda_t)], 
\end{equation}
instead of the counterdiabatic Hamiltonian (\ref{Eq.cd.ham}). 
In this case, only a single projection operator $\hat{P}_n(\lambda_t)$ is transformed as Eq.~(\ref{Eq.ad.trans}), but others $\hat{P}_m(\lambda_t)$ $(m\neq n)$ are not~\cite{Kato1950}.

We introduce the transition rate from the $n$th level to the $m$th level
\begin{align}
p_{nm}(t)&=\|\hat{P}_m(\lambda_t)\hat{U}_D(t)\hat{P}_n(\lambda_0)\hat{U}_D^\dag(t)\hat{P}_m(\lambda_t)\| \nonumber \\
&=\|\hat{P}_m(\lambda_t)\hat{U}_D(t)\hat{P}_n(\lambda_0)\|^2 \nonumber \\
&=\|\hat{P}_m(\lambda_0)\hat{U}_A^\dag(t)\hat{U}_D(t)\hat{P}_n(\lambda_0)\|^2. \label{Eq.trans.rate}
\end{align}
Here, $\|\cdot\|$ is the operator norm, $\|\hat{A}\|\equiv\sqrt{\sup_{|\psi\rangle}\langle\psi|\hat{A}^\dag\hat{A}|\psi\rangle}$. 
Throughout this paper, we only consider bounded operators, and thus $\|A\|=\|A^\dag\|$ holds. 
For non-degenerate eigenstates $\{|n(\lambda_t)\rangle\}$, this quantity is nothing but the transition probability from the $n$th energy eigenstate to the $m$th energy eigenstate 
\begin{equation}
p_{nm}(t)=|\langle m(\lambda_t)|\hat{U}_D(t)|n(\lambda_0)\rangle|^2. 
\end{equation}
For degenerate eigenstates $\{|n,\nu_n(\lambda_t)\rangle\}$, this quantity is the maximum transition probability from an energy eigenstate in the $n$th level to the $m$th level 
\begin{equation}
p_{nm}(t)=\max_{|\psi_n(\lambda_0)\rangle}\sum_{\mu_m}|\langle m,\mu_m(\lambda_t)|\hat{U}_D(t)|\psi_n(\lambda_0)\rangle|^2, 
\end{equation}
where $|\psi_n(\lambda_t)\rangle$ is an $n$th energy eigenstate, which is linear combination of $\{|n,\nu_n(\lambda_t)\rangle\}$.

From the differential equation
\begin{equation}
i\hbar\frac{d}{dt}[\hat{U}_A^\dag(t)\hat{U}_D(t)]=-\hat{U}_A^\dag(t)\hat{H}_\mathrm{cd}(t)\hat{U}_D(t), 
\end{equation}
which follows from Eqs.~(\ref{Eq.Schrodinger.eq}) and (\ref{Eq.tevo.adiabatic}), we obtain
\begin{equation}
\hat{U}_A^\dag(t)\hat{U}_D(t)=1-\frac{1}{i\hbar}\int_0^tdt^\prime\hat{U}_A^\dag(t^\prime)\hat{H}_\mathrm{cd}(t^\prime)\hat{U}_D(t^\prime). 
\label{Eq.pre.uaud}
\end{equation}
Note that $\hat{U}_A^\dag(t)\hat{U}_D(t)$ is unitary, i.e., 
\begin{align}
1=&\hat{U}_A^\dag(t)\hat{U}_D(t)\hat{U}_D^\dag(t)\hat{U}_A(t) \nonumber \\
=&\left(1-\frac{1}{i\hbar}\int_0^tdt^\prime\hat{U}_A^\dag(t^\prime)\hat{H}_\mathrm{cd}(t^\prime)\hat{U}_D(t^\prime)\right) \nonumber \\
&\times\left(1+\frac{1}{i\hbar}\int_0^tdt^\prime\hat{U}_D^\dag(t^\prime)\hat{H}_\mathrm{cd}(t^\prime)\hat{U}_A(t^\prime)\right)
\end{align}
holds, and thus the equality
\begin{align}
&0= \nonumber \\
&\frac{1}{i\hbar}\int_0^tdt^\prime[\hat{U}_D^\dag(t^\prime)\hat{H}_\mathrm{cd}(t^\prime)\hat{U}_A(t^\prime)-\hat{U}_A^\dag(t^\prime)\hat{H}_\mathrm{cd}(t^\prime)\hat{U}_D(t^\prime)] \nonumber \\
&+\frac{1}{\hbar^2}\int_0^tdt^\prime\hat{U}_A^\dag(t^\prime)\hat{H}_\mathrm{cd}(t^\prime)\hat{U}_D(t^\prime)\int_0^tdt^\prime\hat{U}_D^\dag(t^\prime)\hat{H}_\mathrm{cd}(t^\prime)\hat{U}_A(t^\prime), \label{Eq.cond.unitary}
\end{align}
must be satisfied.

For $m\neq n$, by using Eq.~(\ref{Eq.pre.uaud}), the transition rate (\ref{Eq.trans.rate}) becomes
\begin{equation}
p_{nm}(t)=\left\|\frac{1}{\hbar}\int_0^tdt^\prime\hat{P}_m(\lambda_0)\hat{U}_A^\dag(t^\prime)\hat{H}_\mathrm{cd}(t^\prime)\hat{U}_D(t^\prime)\hat{P}_n(\lambda_0)\right\|^2. 
\end{equation}
In general, calculation of time evolution operator $\hat{U}_D(t)$ is hard, and thus we omit it by using the properties of the operator norm as
\begin{align}
p_{nm}(t)&\le\left[\frac{1}{\hbar}\int_0^tdt^\prime\|\hat{P}_m(\lambda_0)\hat{U}_A^\dag(t^\prime)\hat{H}_\mathrm{cd}(t^\prime)\hat{U}_D(t^\prime)\hat{P}_n(\lambda_0)\|\right]^2 \nonumber \\
&\le\left[\frac{1}{\hbar}\int_0^tdt^\prime\|\hat{P}_m(\lambda_0)\hat{U}_A^\dag(t^\prime)\hat{H}_\mathrm{cd}(t^\prime)\|\|\hat{U}_D(t^\prime)\hat{P}_n(\lambda_0)\|\right]^2 \nonumber \\
&=\left[\frac{1}{\hbar}\int_0^tdt^\prime\|\hat{P}_m(\lambda_{t^\prime})\hat{H}_\mathrm{cd}(t^\prime)\|\right]^2 \nonumber \\
&=\left[\frac{1}{\hbar}\int_0^tdt^\prime\|\hat{H}_\mathrm{cd}(t^\prime)\hat{P}_m(\lambda_{t^\prime})\|\right]^2 \nonumber \\
&=\left[\int_0^tdt^\prime\|[1-\hat{P}_m(\lambda_{t^\prime})]\dot{\hat{P}}_m(\lambda_{t^\prime})\|\right]^2. \label{Eq.bound.pre}
\end{align}
Here, we use $\|\hat{P}_n(\lambda_0)\|=1$ for the equality of the third line and Eq.~(\ref{Eq.cd.ham}) for the equality of the last line. 
For non-degenerate eigenstates $\{|n(\lambda_t)\rangle\}$, the integrand of Eq.~(\ref{Eq.bound.pre}) is nothing but the Abelian quantum geometric tensor for the $m$th energy eigenstate
\begin{equation}
\|[1-\hat{P}_m(\lambda_{t})]\dot{\hat{P}}_m(\lambda_{t})\|^2=\langle\dot{m}(\lambda_t)|(1-|m(\lambda_t)\rangle\langle m(\lambda_t)|)|\dot{m}(\lambda_t)\rangle, 
\label{Eq.qgt}
\end{equation}
and for degenerate eigenstates $\{|n,\nu_n(\lambda_t)\rangle\}$, it is the maximum expectation value of the non-Abelian quantum geometric tensor for the $m$th level
\begin{align}
&\|[1-\hat{P}_m(\lambda_{t})]\dot{\hat{P}}_m(\lambda_{t})\|^2 \nonumber \\
=&\max_{|\psi_m(\lambda_t)\rangle}\sum_{\mu_m}\langle\psi_m(\lambda_t)|m,\mu_m(\lambda_t)\rangle\langle\dot{m,\mu_m}(\lambda_t)| \nonumber \\
&\quad\quad\times(1-\sum_{\nu_m}|m,\nu_m(\lambda_t)\rangle\langle m,\nu_m(\lambda_t)|) \nonumber \\
&\quad\quad\times\sum_{\mu_m^\prime}|\dot{m,\mu_m^\prime}(\lambda_t)\rangle\langle m,\mu_m^\prime(\lambda_t)|\psi_m(\lambda_t)\rangle. 
\end{align}
The real part of the quantum geometric tensor gives distance between quantum states with slightly different parameters~\cite{Ma2010}. 
Therefore, singular behavior of the quantum geometric tensor for energy eigenstates suggests the existence of (excited state) quantum phase transitions. 
The above result implies that nonadiabatic transitions are affected by geometry of excited states.

Next, we will consider the remaining rate, i.e., the transition rate (\ref{Eq.trans.rate}) for $m=n$. 
In this case, by using Eqs.~(\ref{Eq.pre.uaud}) and (\ref{Eq.cond.unitary}), we find
\begin{align}
&p_{nn}(t) \nonumber \\
=&\left\|\hat{P}_n(\lambda_0)\left(1-\frac{1}{i\hbar}\int_0^tdt^\prime\hat{U}_A^\dag(t^\prime)\hat{H}_\mathrm{cd}(t^\prime)\hat{U}_D(t^\prime)\right)\hat{P}_n(\lambda_0)\right. \nonumber \\
&\quad\left.\times\left(1+\frac{1}{i\hbar}\int_0^tdt^\prime\hat{U}_D^\dag(t^\prime)\hat{H}_\mathrm{cd}(t^\prime)\hat{U}_A(t^\prime)\right)\hat{P}_n(\lambda_0)\right\| \nonumber \\
=&\left\|\hat{P}_n(\lambda_0)-\frac{1}{\hbar^2}\int_0^tdt^\prime\hat{P}_n(\lambda_0)\hat{U}_A^\dag(t^\prime)\hat{H}_\mathrm{cd}(t^\prime)\hat{U}_D(t^\prime)\right. \nonumber \\
&\quad\times\sum_{\substack{m \\ (m\neq n)}}\hat{P}_m(\lambda_0)\left.\int_0^tdt^\prime\hat{U}_D^\dag(t^\prime)\hat{H}_\mathrm{cd}(t^\prime)\hat{U}_A(t^\prime)\hat{P}_n(\lambda_0)\right\| \nonumber \\
\ge&\|\hat{P}_n(\lambda_0)\| \nonumber \\
&-\frac{1}{\hbar^2}\left\|\int_0^tdt^\prime\hat{P}_n(\lambda_0)\hat{U}_A^\dag(t^\prime)\hat{H}_\mathrm{cd}(t^\prime)\hat{U}_D(t^\prime)\right. \nonumber \\
&\quad\times\sum_{\substack{m \\ (m\neq n)}}\hat{P}_m(\lambda_0)\left.\int_0^tdt^\prime\hat{U}_D^\dag(t^\prime)\hat{H}_\mathrm{cd}(t^\prime)\hat{U}_A(t^\prime)\hat{P}_n(\lambda_0)\right\| \nonumber \\
\ge&1-\left[\frac{1}{\hbar}\int_0^tdt^\prime\|\hat{P}_n(\lambda_{t^\prime})\hat{H}_\mathrm{cd}(t^\prime)\|\right]^2 \nonumber \\
=&1-\left[\frac{1}{\hbar}\int_0^tdt^\prime\|\hat{H}_\mathrm{cd}(t^\prime)\hat{P}_n(\lambda_{t^\prime})\|\right]^2 \nonumber \\
=&1- \left[\int_0^tdt^\prime\|[1-\hat{P}_n(\lambda_{t^\prime})]\dot{\hat{P}}_n(\lambda_{t^\prime})\|\right]^2. \label{Eq.bound.rem}
\end{align}
In the last inequality and the equality, we use similar caluculations to Eq.~(\ref{Eq.bound.pre}).

We can also derive a universal bound for any nonadiabatic transition. 
By using the inequality
\begin{equation}
\|\hat{P}_m(\lambda_t)\hat{H}_\mathrm{cd}(t)\|\le\|\hat{H}_\mathrm{cd}(t)\|, 
\label{Eq.ineq.uni}
\end{equation}
we immediately find a universal bound for the transition rate
\begin{equation}
p_{nm}(t)\le\left[\frac{1}{\hbar}\int_0^tdt^\prime\|\hat{H}_\mathrm{cd}(t^\prime)\|\right]^2, 
\label{Eq.bound.CD}
\end{equation}
and similarly a universal bound for the remaining rate
\begin{equation}
p_{nn}(t)\ge1-\left[\frac{1}{\hbar}\int_0^tdt^\prime\|\hat{H}_\mathrm{cd}(t^\prime)\|\right]^2.  
\label{Eq.bound.CD2}
\end{equation}
The small counterdiabatic Hamiltonian implies that small perturbation is enough to be adiabatic, and thus this bound is a physically natural. 
Note that this bound is related to the energy cost of counterdiabatic driving~\cite{Santos2015,Coulamy2016}.

\section{Adiabatic perturbation theory viewpoint}
We discuss our results from the viewpoint of the adiabatic perturbation theory~\cite{Polkovnikov2005,Polkovnikov2011,Kolodrubetz2017}. 
A quantum quench at time $t$ changes the $m$th projection operator as
\begin{equation}
\hat{P}_m(\lambda_{t+\delta t})\approx\hat{P}_m(\lambda_t)+\delta t\dot{\hat{P}}_m(\lambda_t),
\end{equation}
where $\delta t$ is a small time interval. 
Then, we find that a quantum quench causes nonadiabatic transitions from the $n$th level to the $m$th level as
\begin{equation}
\hat{P}_m(\lambda_{t+\delta t})\hat{P}_n(\lambda_t)\hat{P}_m(\lambda_{t+\delta t})\approx\delta t^2\dot{\hat{P}}_m(\lambda_t)\hat{P}_n(\lambda_t)\dot{\hat{P}}_m(\lambda_t),
\end{equation}
that is, the instantaneous transition rate at time $t$ for a quantum quench is given by
\begin{align}
\Delta p_{nm}^\mathrm{APT}(t)&=\|\hat{P}_m(\lambda_{t+\delta t})\hat{P}_n(\lambda_t)\hat{P}_m(\lambda_{t+\delta t})\| \nonumber \\
&\approx\delta t^2\|\hat{P}_n(\lambda_t)\dot{\hat{P}}_m(\lambda_t)\|^2. 
\end{align}

However, starting from $t=0$, a state is distributed over various levels because of successive nonadiabatic transitions. 
Therefore, to consider the quantity (\ref{Eq.trans.rate}), we rather sum up with respect to $n$ except for $m$ to find the amount of nonadiabatic transitions to $m$th level at time $t$, i.e., 
\begin{align}
&\sum_{\substack{n \\ (n\neq m)}}\hat{P}_m(\lambda_{t+\delta t})\hat{P}_n(\lambda_t)\hat{P}_m(\lambda_{t+\delta t}) \nonumber \\
&\approx\delta t^2\dot{\hat{P}}_m(\lambda_t)[1-\hat{P}_m(\lambda_t)]\dot{\hat{P}}_m(\lambda_t). 
\end{align}
Then we find the instantaneous transition rate to $m$th level at time $t$
\begin{align}
\Delta p_{m}^\mathrm{APT}(t)&=\left\|\sum_{\substack{n \\ (n\neq m)}}\hat{P}_m(\lambda_{t+\delta t})\hat{P}_n(\lambda_t)\hat{P}_m(\lambda_{t+\delta t})\right\| \nonumber \\
&\approx\delta t^2\|[1-\hat{P}_m(\lambda_t)]\dot{\hat{P}}_m(\lambda_t)\|^2. 
\end{align}
Finally we find similar quantity to the bound (\ref{Eq.bound.pre}) by using the adiabatic perturbation theory, and thus our result is consistent with an approximate calculation by the adiabatic perturbation theory. 
A calculate of the total transition rate (\ref{Eq.trans.rate}) by using the adiabatic perturbation theory is usually difficult because we have to accumulate all the paths of instantaneous nonadiabatic transitions resulting in the $m$th level.

\section{\label{Sec.QSL}Quantum speed limit viewpoint}
In the quantum speed limit~\cite{Mandelstam1945,Pfeifer1995,Margolus1998,Deffner2017}, we introduce the distance between two quantum states, $\hat{\rho}_i$ and $\hat{\rho}_f$, by the Bures angle~\cite{Kakutani1948,Bures1969}
\begin{equation}
L(\hat{\rho}_i,\hat{\rho}_f)=\arccos\sqrt{F(\hat{\rho}_i,\hat{\rho}_f)},
\end{equation}
where $F(\hat{\rho}_i,\hat{\rho}_f)$ is the Uhlmann fidelity~\cite{Uhlmann1976}
\begin{equation}
F(\hat{\rho}_i,\hat{\rho}_f)=\left(\mathrm{Tr}\sqrt{\sqrt{\hat{\rho}_i}\hat{\rho}_f\sqrt{\hat{\rho}_i}}\right)^2. 
\end{equation}
We consider time evolution governed by
\begin{equation}
i\hbar\dot{\hat{\rho}}(t)=[\hat{H}(\lambda_t),\hat{\rho}(t)], 
\end{equation}
satisfying $\hat{\rho}(0)=\hat{\rho}_i$ and $\hat{\rho}(\tau)=\hat{\rho}_f$. 
It can be shown that this distance is bounded as
\begin{equation}
L(\hat{\rho}_i,\hat{\rho}_f)\le\frac{1}{\hbar}\int_0^\tau dt(\Delta\hat{H})_{\hat{\rho}_t},
\label{Eq.speed.limit}
\end{equation}
where $(\Delta\hat{H})_{\hat{\rho}_t}$ is the standard deviation of the Hamiltonian $\hat{H}(\lambda_t)$ with a state $\hat{\rho}(t)$~\cite{Uhlmann1992,Deffner2013}. 
This bound means that a quantum process between large distant states requires large energy cost or long operation time.

Here we consider a state in the $m$th level $\hat{\rho}_m(0)=\hat{P}_m(\lambda_0)\hat{\rho}_m(0)\hat{P}_m(\lambda_0)$ and its adiabatic transformation $\hat{\rho}_m(t)=\hat{U}_A(t)\hat{\rho}_m(0)\hat{U}_A^\dag(t)$. 
For this adiabatic transformation, the quantum speed limit becomes~\cite{Funo2017}
\begin{equation}
L(\hat{\rho}_m(0),\hat{\rho}_m(t))\le\frac{1}{\hbar}\int_0^tdt^\prime(\Delta\hat{H}_\mathrm{cd})_{\hat{\rho}_m}. 
\end{equation}
Here, following the inequality holds: 
\begin{equation}
\begin{aligned}
\frac{1}{\hbar}(\Delta\hat{H}_\mathrm{cd})_{\hat{\rho}_m}=&\frac{1}{\hbar}\sqrt{\mathrm{Tr}\{[\hat{H}_\mathrm{cd}(t)]^2\hat{\rho}_m(t)\}-\{\mathrm{Tr}[\hat{H}_\mathrm{cd}(t)\hat{\rho}_m(t)]\}^2} \\
=&\frac{1}{\hbar}\sqrt{\mathrm{Tr}\{[\hat{H}_\mathrm{cd}(t)]^2\hat{\rho}_m(t)\}} \\
\le&\frac{1}{\hbar}\|\hat{H}_\mathrm{cd}(t)\hat{P}_m(\lambda_t)\| \\
=&\|[1-\hat{P}_m(\lambda_t)]\dot{\hat{P}}_m(\lambda_t)\|. 
\end{aligned}
\end{equation}
In particular, nondegenerate eigenstates satisfies equality. 
Finally, the following bound exists: 
\begin{equation}
[L(\hat{\rho}_m(0),\hat{\rho}_m(t))]^2\le\left[\int_0^tdt^\prime\|[1-\hat{P}_m(\lambda_{t^\prime})]\dot{\hat{P}}_m(\lambda_{t^\prime})\|\right]^2. 
\label{Eq.bound.qsl}
\end{equation}
Therefore, nonadiabatic transitions to the $m$th level caused by the dynamical transformation shares the identical upper bound with the quantum speed limit of the adiabatic transformation for the $m$th level.

Suppose that the bound (\ref{Eq.bound.pre}) [(\ref{Eq.bound.qsl})] is small, which is of interest. 
From the viewpoint of the quantum speed limit, it implies that the adiabatic transformation from $\hat{\rho}_m(0)$ to $\hat{\rho}_m(t)$ can be implemented without requiring large energy cost or long operation time. 
At the same time, the bound (\ref{Eq.bound.pre}) implies that the amount of nonadiabatic transitions to $m$th level is small. 
This result can be interpreted as follows: 
The distance between $\hat{\rho}_m(0)$ and $\hat{\rho}_m(t)$ is small. 
From the definition, the distance between $\hat{\rho}_m(0)$ and $\hat{\rho}_n(0)$ is large. 
Therefore, the distance between $\hat{\rho}_n(0)$ and $\hat{\rho}_m(t)$ is large, and thus nonadiabatic transitions do not take place that much.

\section{Example}

As an example, we consider a two-level system
\begin{equation}
\hat{H}(\lambda_t)=h_t^x\hat{X}+h_t^y\hat{Y}+h_t^z\hat{Z},
\end{equation}
where $\lambda_t\equiv\bm{h}_t=(h_t^x,h_t^y,h_t^z)$ is the time-dependent parameters and $\hat{X}$, $\hat{Y}$, and $\hat{Z}$ are the Pauli matrices. 
In this section, we set $\hbar=1$. 
For this system, the square root of Eq. (\ref{Eq.qgt}) is given by
\begin{equation}
\|[1-\hat{P}_\pm(\lambda_t)]\dot{\hat{P}}_\pm(\lambda_t)\|=\frac{|\bm{h}_t\times\dot{\bm{h}}_t|}{2|\bm{h}_t|^2},
\end{equation}
where $\hat{P}_-(\lambda_t)$ [$\hat{P}_+(\lambda_t)$] is the projection operator for the ground (excited) state. 
Note that the equality of Eq.~(\ref{Eq.ineq.uni}) is satisfied for this system, and thus the bound (\ref{Eq.bound.pre}) [(\ref{Eq.bound.rem})] is identical to the universal bound (\ref{Eq.bound.CD}) [(\ref{Eq.bound.CD2})]. 
By rewriting the time-dependent parameters in the spherical coordinates as $\bm{h}_t=|\bm{h}_t|(\sin\theta_t\cos\phi_t,\sin\theta_t\sin\phi_t,\cos\theta_t)$, we find
\begin{equation}
\int_0^tdt^\prime\|[1-\hat{P}_\pm(\lambda_t)]\dot{\hat{P}}_\pm(\lambda_t)\|=\frac{1}{2}\int_0^tdt^\prime\sqrt{\dot{\theta}_{t^\prime}^2+\dot{\phi}_{t^\prime}^2\sin^2\theta_{t^\prime}}. 
\label{Eq.traj.leng}
\end{equation}
This quantity is nothing but the half length of the trajectory of $\bm{h}(\lambda_t)/|\bm{h}(\lambda_t)|$, i.e., the trajectory of the energy eigenstate on the Bloch sphere. 
The amount of nonadiabatic transitions is bounded by the square of Eq.~(\ref{Eq.traj.leng}).

For concreteness, we consider quantum annealing in this system, i.e., we set $\bm{h}_0=(h_0^x,0,0)$ and $\bm{h}_T=(0,0,h_T^z)$, where $T$ is the annealing time, and the parameters $h_0^x$ and $h_T^z$ are arbitrary non-zero values. 
Here we set $h_0^x<0$ and $h_T^z<0$. 
In this case, Eq.~(\ref{Eq.traj.leng}) is the half length of the trajectory on the Bloch sphere from $(1,0,0)$ to $(0,0,1)$. 
Therefore, its minimum value is $\pi/4$, and thus the transition rate of the optimal case is bounded as $p_{-+}(T)\le(\pi/4)^2\approx0.62$. 
We note that in the quench limit $T\to0$ the transition rate is $p_{-+}(T)|_{T\to0}=0.5$.

Finally, we discuss possible application of this example to other general systems. 
We consider the general Hamiltonian (\ref{Eq.ham}) in the non-degenerate case and the Schr\"odinger equation
\begin{equation}
i\frac{\partial}{\partial t}|\Psi(t)\rangle=\hat{H}(\lambda_t)|\Psi(t)\rangle. 
\label{Eq.schro1}
\end{equation}
By expanding the state $|\Psi(t)\rangle$ by the energy eigenstates $\{|n(\lambda_t)\rangle\}$ as $|\Psi(t)\rangle=\sum_n\psi_n(t)|n(\lambda_t)\rangle$, the Schr\"odinger equation (\ref{Eq.schro1}) is rewritten as~\cite{Kolodrubetz2017}
\begin{equation}
i\frac{\partial}{\partial t}\bm{\psi}(t)=\hat{\tilde{H}}(t)\bm{\psi}(t),
\end{equation}
where $\bm{\psi}(t)={}^t(\psi_0(t),\psi_1(t),\dots)$ and 
\begin{equation}
\hat{\tilde{H}}(t)=
\begin{pmatrix}
E_0(\lambda_t) & -i\langle0(\lambda_t)|\dot{1}(\lambda_t)\rangle & \cdots \\
-i\langle1(\lambda_t)|\dot{0}(\lambda_t)\rangle & E_1(\lambda_t) & \cdots \\
\vdots &\vdots & \ddots
\end{pmatrix}. 
\label{Eq.manybody.ham}
\end{equation}
We assume that dynamics is almost adiabatic, and thus we only focus on the ground state and the first excited state, i.e., we consider the projected two-level system~\cite{Chen2020}. 
In this case, the Hamiltonian (\ref{Eq.manybody.ham}) is approximately given by
\begin{equation}
\begin{aligned}
\hat{\tilde{H}}(t)\approx&\mathrm{Im}\langle0(\lambda_t)|\dot{1}(\lambda_t)\rangle\hat{X}+\mathrm{Re}\langle0(\lambda_t)|\dot{1}(\lambda_t)\rangle\hat{Y} \\
&+\frac{E_0(\lambda_t)-E_1(\lambda_t)}{2}\hat{Z}, 
\end{aligned}
\end{equation}
where we neglect the term proportional to the identity operator. 
Therefore, we can apply the above discussion to general systems if we know the ground state and the first excited state. 
We leave detailed analysis of this method for the future work.

\section{Summary and discussion}
In this paper, we discussed the bounds for nonadiabatic transitions from the viewpoints of the adiabatic perturbation theory and the quantum speed limit. 
From Eq.~(\ref{Eq.bound.pre}), we found that the amount of nonadiabatic transitions from the $n$th level to the $m$th level is bounded by the function of the quantum geometric tensor for the $m$th level. 
This bound does not depend on $n$, but it might be natural because we have to accumulate all the paths of instantaneous nonadiabatic transitions as discussed from the viewpoint of the adiabatic perturbation theory. 
As the result of accumulation, this bound gives the worst case value and thus it may overestimate the amount of nonadiabatic transitions. 
The viewpoint of the quantum speed limit provides us with nontrivial relationship between the quantum speed limit of the adiabatic transformation for the $m$th level and nonadiabatic transitions to the $m$th level caused by the dynamical transformation. 
We also derived the universal bound for any nonadiabatic transition written in terms of the counterdiabatic Hamiltonian.

In quantum adiabatic brachistochrone, we reduce nonadiabatic transitions by minimizing cost functions related to adiabaticity~\cite{Rezakhani2009,Rezakhani2010,Takahashi2019}. 
We note that the bound (\ref{Eq.bound.pre}) for nondegenerate eigenstates is the square of a cost function used in Ref.~\cite{Rezakhani2010}. 
We have many choice for cost functions in quantum adiabatic brachistochrone~\cite{Rezakhani2009,Rezakhani2010,Takahashi2019}, but the present result suggests that the cost function used in Ref.~\cite{Rezakhani2010} is physically natural. 
Note that physically natural choice does not imply the best choice. 
From this relationship, we can easily find an optimal schedule of parameter $\lambda_t$ minimizing the transition rate $p_{nm}(t)$ by using quantum adiabatic brachistochrone. 
Success of quantum adiabatic brachistochrone in adiabatic state preparation reported in literature~\cite{Takahashi2019,Hatomura2019} also supports effectiveness of the bound~(\ref{Eq.bound.pre}).

Next, we will explain that the universal bound~(\ref{Eq.bound.CD}) may be more useful than the bound~(\ref{Eq.bound.pre}) in some situations, while the universal bound~(\ref{Eq.bound.CD}) is looser than the bound~(\ref{Eq.bound.pre}). 
Both of the bounds~(\ref{Eq.bound.pre}) and (\ref{Eq.bound.CD}) in principle requires knowledge of energy eigenstates, and this requirement makes it hard to obtain these bounds for general quantum systems. 
However, it is possible to construct counterdiabatic terms without knowing energy eigenstates of a system by using a variational approach~\cite{Sels2017}, i.e., we can estimate the value of the bound~(\ref{Eq.bound.CD}).  
Furthermore, this can be systematically performed by calculating the nested commutators $[\hat{H}(\lambda_t),[\hat{H}(\lambda_t),[\cdots,[\hat{H}(\lambda_t),\dot{\hat{H}}(\lambda_t)]]\cdots]$~\cite{Claeys2019}.

Finally, we mention related works. 
In Refs.~\cite{Lychkovskiy2017} and \cite{Suzuki2019}, deviation of a dynamical state from an adiabatic state was quantified by using the quantum speed limit. 
Their discussion is related to the remaining rate (\ref{Eq.bound.rem}) in the present paper, i.e., they considered the $nn$-component of the transition probability (\ref{Eq.trans.rate}). 
Our results can be understood as the extension of it to other components. 
We also provided physical meaning of the bound from the viewpoint of the adiabatic perturbation theory and the quantum speed limit.

\bibliography{nonadbib}

\end{document}